\documentclass[reprint,amsmath,amssymb,aps,prx]{revtex4-2}
\usepackage{xcolor}
\usepackage{graphicx}
\usepackage[utf8]{inputenc}
\usepackage{physics}
\usepackage{amsmath}
\usepackage{amsfonts}
\usepackage{dcolumn}
\usepackage{bm}
\usepackage{wrapfig}
\usepackage[normalem]{ulem}
\usepackage[colorlinks=true, hyperindex, breaklinks, linkcolor=blue, urlcolor=blue, citecolor=blue]{hyperref}
\usepackage{setspace}
\usepackage[T1]{fontenc}
\usepackage{fancyhdr}
\usepackage[separate-uncertainty=true,multi-part-units=single]{siunitx}

\fancypagestyle{footertext}{
    \fancyhf{} 
    \fancyfoot[C]{\thepage} 
}
\DeclareUnicodeCharacter{2009}{\,}

\makeatletter
\def\maketitle{
	\@author@finish
	\title@column\titleblock@produce
	\suppressfloats[t]}
\makeatother 

\usepackage{xpatch}

\makeatletter
\patchcmd{\@ssect@ltx}
{\addcontentsline{toc}{#1}{\protect\numberline{}#8}}
{}
{}
{}
\makeatother

\begin{document}

\title{Superconducting cavity qubit with tens of milliseconds single-photon coherence time}
\author{Ofir Milul}
\thanks{These authors contributed equally to this work.}
\author{Barkay Guttel}
\thanks{These authors contributed equally to this work.}
\author{Uri Goldblatt}
\author{Sergey Hazanov}
\author{Lalit M. Joshi}
\author{Daniel Chausovsky}
\author{Nitzan Kahn}
\author{Engin \c{C}ifty\"{u}rek}
\author{Fabien Lafont}
\author{Serge Rosenblum}

\affiliation{
\vspace{3pt}Department of Condensed Matter Physics, Weizmann Institute of Science, Rehovot, Israel
}


\begin{abstract}
Storing quantum information for an extended period of time is essential for running quantum algorithms with low errors. Currently, superconducting quantum memories have coherence times of a few milliseconds, and surpassing this performance has remained an outstanding challenge.  In this work, we report a single-photon qubit encoded in a novel superconducting cavity with a coherence time of 34 ms, representing an order of magnitude improvement compared to previous demonstrations. We use this long-lived quantum memory to store a Schr\"odinger cat state with a record size of 1024 photons, indicating the cavity's potential for bosonic quantum error correction. 

\end{abstract}

\maketitle

\section{Introduction}
Superconducting qubits have emerged as a leading technology for quantum computing due to their scalability and high gate fidelities, which exceed the threshold for fault tolerance \cite{Barends2014SuperconductingTolerance}.
However, further extending the coherence times of these qubits remains a crucial area of research. Longer coherence times can support higher gate fidelities, thereby reducing the hardware overhead for achieving fault tolerance. Additionally, high-coherence qubits can serve as quantum memories \cite{Mariantoni2011ImplementingCircuits}, enabling the efficient execution of sequential quantum algorithms \cite{Gouzien2021FactoringMemory,Thaker2006QuantumComputing} and playing a vital role in quantum networks by storing entanglement between distant nodes \cite{Flurin2015SuperconductingRadiation}.

To date, single-photon states of three-dimensional microwave cavities are among the longest-lived qubits measured in any superconducting device, with coherence times exceeding $2\,$ms \cite{Kjaergaard2020SuperconductingPlay,Rosenblum2018Fault-tolerantError,Chakram2021SeamlessElectrodynamics}. The single-photon lifetimes of these quantum memories are currently limited by lossy oxides on the cavity surface \cite{Romanenko2017UnderstandingAmplitudes, Kudra2020HighCavities}. While eliminating these surface oxides has resulted in cavities with quality factors exceeding $Q\sim10^{10}$ at low fields \cite{Romanenko2020Three-DimensionalS,Posen2020UltralowCavities,Heidler2021Non-MarkovianMilliseconds}, this remains to be translated into corresponding improvements in qubit coherence times. 
A major challenge towards achieving this goal is the need to couple these cavities to error-prone transmons \cite{Paik2011ObservationArchitecture,Reagor2013ReachingCavities} for qubit encoding, manipulation, and decoding. This coupling introduces new loss channels to the cavity mode, thereby reducing the single-photon lifetime of the cavity \cite{Wang2015SurfaceQubits,Read2022PrecisionSensitivity,Reagor2016QuantumQED}. Additionally, transmon heating events cause the cavity frequency to fluctuate, degrading the coherence time of the cavity \cite{Sears2012PhotonQED,Rigetti2012SuperconductingMs}. Finally, the coupling to the transmon leads to an undesired Kerr nonlinearity of the cavity \cite{Kirchmair2013ObservationEffect}.
\begin{figure}[b!]
        \vspace{-10pt}
\includegraphics[scale=1.57]{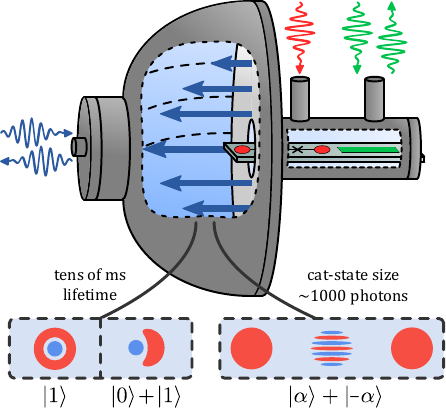}
    \vspace{-4pt}
    \caption{Illustration of the superconducting cavity qubit. The high-purity niobium cavity (top) consists of a half-elliptical part and a flat part, which are electron-beam welded together. We use the fundamental $\textrm{TM}_{010}$ mode of this cavity, whose maximum electric field (straight blue arrows) is near the opening of the waveguide housing the transmon chip. This design allows for effective coupling to the transmon (red pads) with minimal chip protrusion into the cavity. The transmon is coupled to an on-chip stripline resonator (green rectangle) used for transmon and cavity measurements (green arrows). Classical ring-down measurements of the cavity (wavy blue arrows) are performed through an undercoupled cavity pin. The Wigner distributions of single-photon qubit states (bottom left) and a Schr\"odinger cat state (bottom right) illustrate the stored states demonstrated in this work.}
  \label{fig:setup}
\end{figure}

\indent In this work, we overcome these challenges and demonstrate a long-lived single-photon qubit using a novel superconducting microwave cavity. By coupling the ancilla transmon very weakly to the cavity and using strong parametric drives to compensate for lower interaction rates \cite{Rosenblum2018Fault-tolerantError,Campagne-Ibarcq2020QuantumOscillator,Eickbusch2022FastQubit,Diringer2023ConditionalQubit}, we are able to mitigate transmon-induced cavity errors and nonlinear effects. Our experimental results show a single-photon relaxation time of $T_1^c=25.6\pm 0.2\,$ms and a coherence time of $T_2^c=34\pm 1\,$ms, exceeding previous demonstrations by an order of magnitude. This novel cavity not only serves as a quantum memory with coherence times far exceeding those of on-chip processing qubits, but also opens up new possibilities for bosonic quantum error correction \cite{Joshi2021QuantumQED,Cai2021BosonicCircuits,Grimsmo2021QuantumCode} with higher photon populations. To demonstrate this potential, we prepare and characterize a Schr\"odinger cat state \cite{Deleglise2008ReconstructionDecoherence} with a size of 1024 photons, an increase of an order of magnitude over previous demonstrations \cite{Vlastakis2013DeterministicallyStates,Wang2016ABoxes}.

\section{cavity characterization}
The superconducting niobium cavity developed in this work (see Fig. \ref{fig:setup}) was designed to minimize photon loss caused by exposure to lossy surfaces \cite{Reagor2013ReachingCavities} (see Appendix \ref{Sec:CavityLossChannels}). The cavity is made of two high-purity niobium parts that were electron-beam welded \cite{Aune2000SuperconductingCavities} to eliminate seam loss, and polished to remove damaged surface layers and reduce surface roughness (see Appendix \ref{Sec:Cavity manufacturing and surface preparation}). We chose a half-elliptical geometry for the cavity to maximize the electric field at the center of the flat surface, where a transmon chip housed in a narrow waveguide extends into the cavity by $\sim1$ mm. This design allows for sufficient coupling between the cavity and the transmon while minimizing chip-induced losses \cite{Wang2015SurfaceQubits}. 

\begin{figure}[h!]
\centering
  \includegraphics[width=0.45\textwidth]{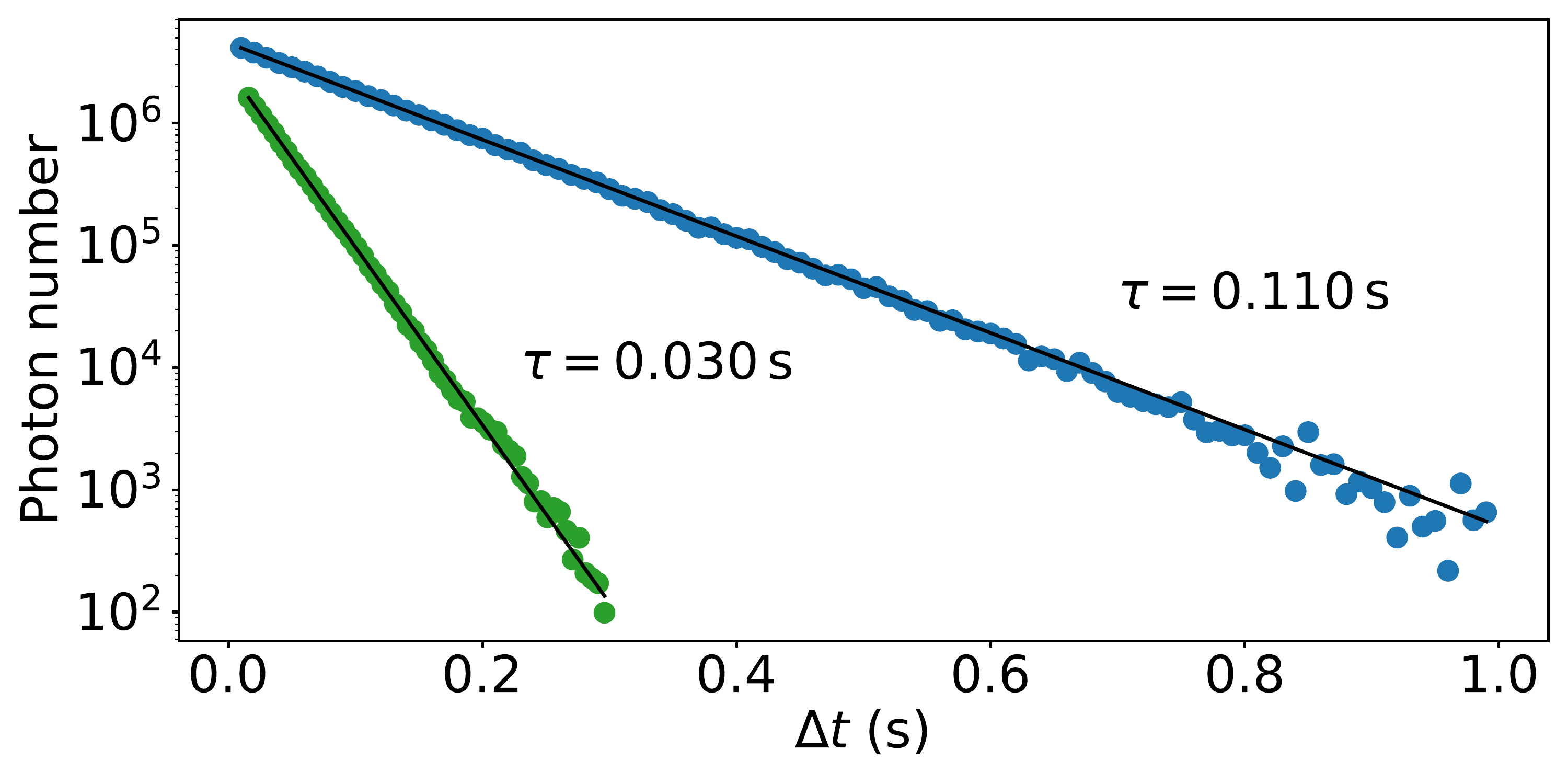}
  \vspace{-4pt}
    \caption{Classical ring-down measurements of the bare cavity. Blue markers show data collected shortly after the initial etching process, with a measured decay time of $\tau=0.110 \,$s. A subsequent degradation of the cavity surface reduced the decay time to $30 \,$ms (green markers).}
  \label{fig:ringdown}
\end{figure}

Before integrating the transmon chip, we characterized the performance of the bare cavity. We conducted a ring-down measurement of the fundamental cavity mode with resonance frequency $\omega_c/2\pi=4.3 \,$GHz. This measurement, performed at \SI{10}{mK}, yielded a bare cavity energy decay time of $0.11\,$s (see Fig. \ref{fig:ringdown}), corresponding to a loaded quality factor of $Q_0=3\times10^9$. The measured external quality factor is $Q_0^\textrm{ext} = 1.3 \times 10^{10}$, implying an intrinsic cavity decay time of \SI{0.14}{s}. This decay time approaches those reported in recent studies of heat-treated niobium cavities, despite the absence of a vacuum annealing step in our process \cite{Romanenko2020Three-DimensionalS,Posen2020UltralowCavities}. The decay time remained stable for multiple cooldowns and under prolonged atmospheric exposure, but ultimately degraded to $30 \,$ms (see Fig. \ref{fig:ringdown}). Subsequent data were collected at this state, which persisted despite repeated etching steps.

\section{Quantum Memory performance}
Next, we inserted a transmon chip \cite{Axline2016AnDevices} into the cavity to enable its use as a quantum memory (see Appendix \ref{Sec:Transmon chip integration}). The aluminum transmon has a resonance frequency of $\omega_q/2\pi=3.1\,$ GHz, energy and coherence lifetimes of $T_1^q=\SI{110}{\micro\second}$, $T_{2 \! \hspace{0.8pt} E}^q=\SI{80}{\micro\second}$ ($T_2^q=\SI{16}{\micro\second}$ without echo), and an anharmonicity of $K_q/2\pi=146 \,$MHz. The transmon and the cavity are dispersively coupled at a rate $\chi/2\pi=42$ kHz. To overcome this weak dispersive interaction \cite{Heeres2017ImplementingOscillator}, we use a parametric sideband interaction \cite{Zeytinoglu2015Microwave-inducedElectrodynamics,Pechal2014Microwave-controlledElectrodynamics,Rosenblum2018ACavities} to map a qubit encoded in the transmon to the cavity with high fidelity. 

A sideband drive at a frequency $2\omega_{q}-K_q-\omega_c$ induces oscillations between  $\ket{0}\ket{f}$ and $\ket{1}\ket{g}$, with $\ket{0}$, $\ket{1}$ referring to vacuum and the single-photon Fock state in the cavity, and $\ket{g}$, $\ket{f}$ to the transmon ground and second excited states.
We obtain a sideband Rabi oscillation rate $\Omega/2\pi = 476\, $kHz, an order of magnitude faster than the limit imposed by $\chi$. 

\noindent The encoding process starts by initializing the transmon in an arbitrary superposition state $\ket{0}(a\ket{g}+b\ket{f})$ and applying the sideband drive for a time $t_\textrm{p}=\pi/\Omega=\SI{1.05}{\micro\second}$. This duration is significantly shorter than the relevant transmon coherence times ($T_2^{g \! \hspace{0.8pt} f}=\SI{45}{\micro\second}$, $T_1^f=\SI{50}{\micro\second}$), resulting in the encoding of a high-fidelity single-photon qubit state $(a\ket{0}+b\ket{1})\ket{g}$ in the cavity. 

After the encoding step, the system idles for the desired storage period. To retrieve the stored state, we apply the sideband pulse a second time. This maps the stored state back onto the transmon, which is then measured.
To prevent excessive relaxation times between experiments, we use a reset drive that couples the cavity to the lossy readout resonator \cite{Pfaff2017ControlledMemory,Wang2016ABoxes}, emptying the cavity within $\sim 2\,$ms (see Appendix \ref{Sec:HamiltonianParameters}).

\begin{figure}[t!]
    \centering
  \includegraphics[width=0.45\textwidth]{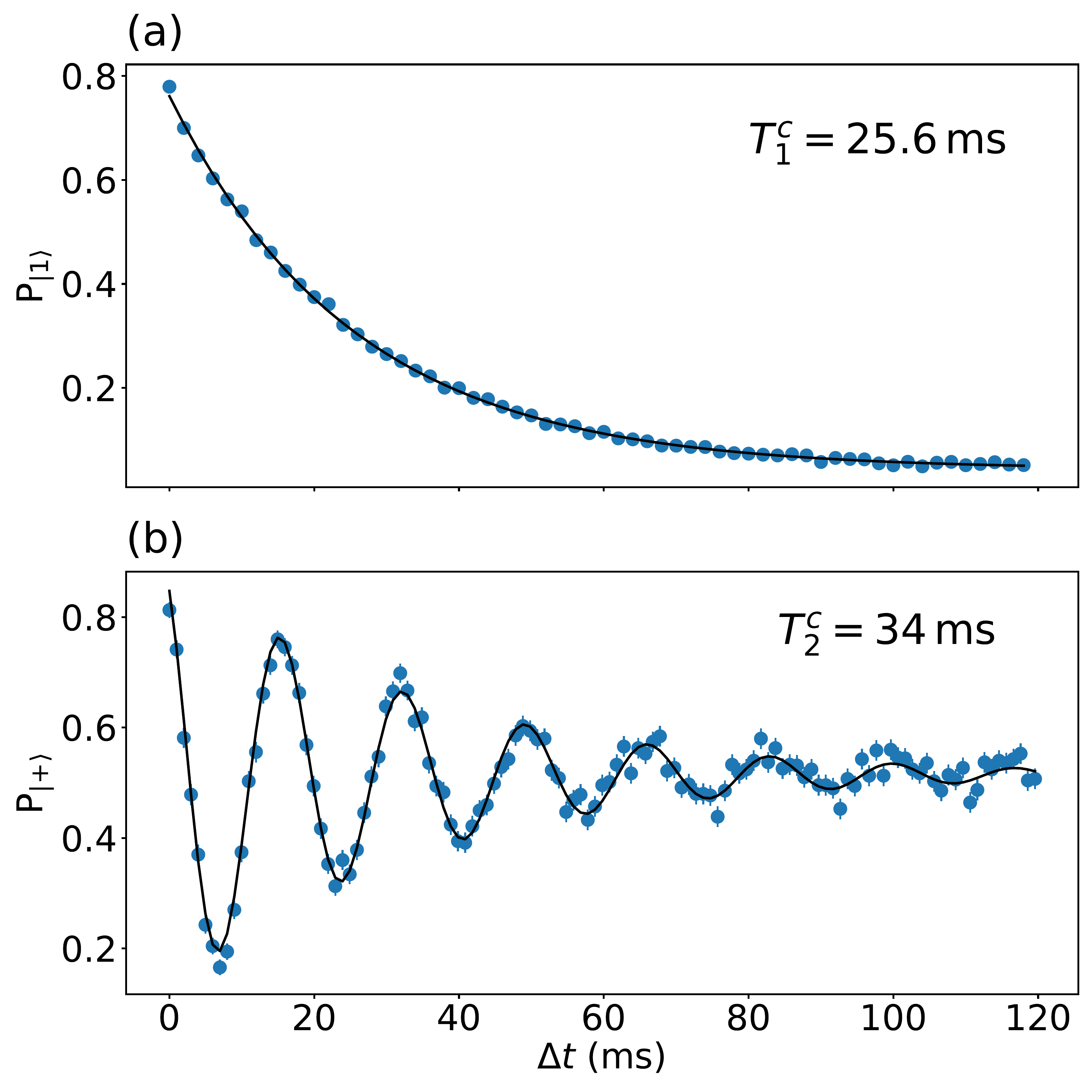}
  \vspace{-4pt}
  \caption{Lifetime of the quantum memory. (\textbf{a}), The single-photon lifetime is obtained by preparing the cavity in a single-photon Fock state $\ket{1}$ and measuring the single-photon probability as a function of the time delay $\Delta t$ (error bars are smaller than the data points). The solid line represents an exponential fit with a characteristic time of $T_1^c = 25.6 \pm 0.2\,$ms.
  (\textbf{b}), The coherence time of the quantum memory is obtained by populating the cavity with the state $\ket{+}=(\ket{0}+\ket{1})/\sqrt{2}$ and measuring the probability of remaining in this state. The sinusoidal exponential fit (solid line) yields a coherence time of $T_2^c = 34 \pm 1\,$ms.}
  \label{fig:T1_T2_plot}
\end{figure}

We characterize the energy lifetime of the quantum memory by preparing a single-photon Fock state in the cavity and measuring its decay time.
We observe a single-photon lifetime of $T_1^c=25.6\pm 0.2\,$ms (see Fig. \ref{fig:T1_T2_plot}(a)). Importantly, these results remained consistent across multiple cooldowns using various cavity etch procedures and different transmon chips (see Appendix \ref{Sec:ThermalPopulationDephasing}). The transmon's contribution to the cavity lifetime through the inverse Purcell effect is estimated \cite{Blais2021CircuitElectrodynamics} to be 
$\frac{K_q}{\chi}T_{2 \! \hspace{0.8pt} E}^q = 278\,$ms. This expression accounts for both transmon relaxation and pure dephasing as factors contributing to the loss of cavity photons (see Appendix \ref{Sec:CavityLossChannels}). As a result, transmon-induced losses are not anticipated to significantly affect the cavity's single-photon lifetime. This observation is supported by the fact that the measured cavity lifetime experiences only a minor reduction when the chip is present.

To determine the coherence time of the quantum memory, we prepare the cavity in a superposition state $(\ket{0}+\ket{1})/\sqrt{2}$ and measure its survival time, yielding $T_2^c=34 \pm 1\,$ms (see Fig. \ref{fig:T1_T2_plot}(b)). We can quantify the different mechanisms contributing to the cavity decoherence using $1/T_2^c = 1/2T_1^c+ 1/T_\uparrow^q + 1/T_\phi^c$. In this expression, $T_\uparrow^q$ is the average time between thermal transmon excitations, which typically cause the cavity to dephase (see Appendix \ref{Sec:ThermalPopulationDephasing}), and $T_\phi^c$ is the dephasing time of the cavity due to other mechanisms. The main contribution to the coherence time $T_2^c$ is photon loss, followed by transmon heating. Using the independently measured average transmon population \cite{Sears2012PhotonQED,Bertet2005DephasingNoise,Geerlings2013DemonstratingQubit,Jin2015ThermalQubit} $\bar{n}_\textrm{th}^q=(1.2\pm0.2)\times10^{-3}$, we deduce that the transmon heating time is $T_\uparrow^q \approx T_1^q/\bar{n}_\textrm{th}^q= 92\pm15\,$ms. Any remaining cavity dephasing mechanism, such as intrinsic cavity dephasing, can only account for a small fraction of the total decoherence, with $T_\phi^c > 0.5\,$s. 
\newline
\indent Finally, we established a lower bound of $T_\uparrow^c \approx T_1^c/\bar{n}_\textrm{th}^c>5\,$s  \cite{Rigetti2012SuperconductingMs} on the cavity heating time. This bound was obtained by applying the sideband drive without initializing the transmon in $\ket{f}$. The absence of sideband oscillations then indicates a thermal cavity photon population of $\bar{n}^c_\textrm{th}<0.5\%$.

\section{Encoding Schr\"odinger cat states}
A key advantage of superconducting cavities is their ability to store quantum information redundantly in multiphoton states, thereby allowing bosonic quantum error correction \cite{Ofek2016ExtendingCircuits,Cai2021BosonicCircuits, Joshi2021QuantumQED, Grimsmo2020QuantumCodes}. However, as the size of the encoded multiphoton states increases, so does the rate of photon loss events. Cavities with low photon loss rates can therefore be beneficial for implementing bosonic quantum error correction schemes with high photon populations \cite{Berdou2023OneOscillator,Regent2022High-performanceOperations,Grimm2020StabilizationQubit}. 
 
To demonstrate the potential of our quantum memory for bosonic error correction, we prepare Schr\"odinger cat states $\mathcal{N}(\ket{\alpha}\pm\ket{-\alpha})$ by applying a parity measurement to coherent states $\ket{\alpha}$ with mean photon number $|\alpha|^2$ \cite{Sun2014TrackingMeasurements} (see Appendix \ref{Sec:Parity_measurement_with_detuned_drive}). The normalization constant $\mathcal{N}$ approaches $1/\sqrt{2}$ as the mean photon number increases.

\begin{figure}[b]
  \includegraphics[width=0.45\textwidth]{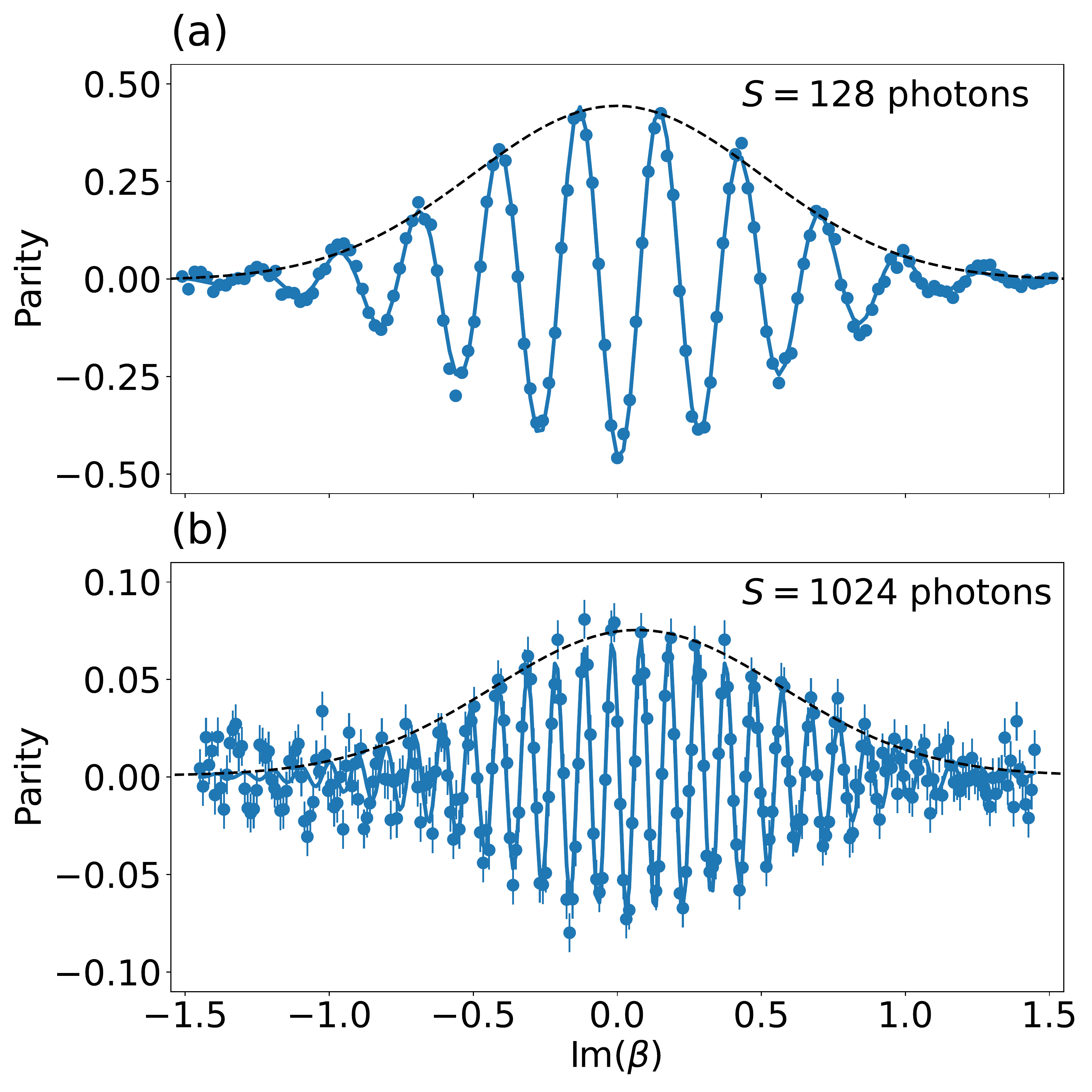}
  \caption{Cut along the imaginary axis of the Wigner phase-space distributions for Schr\"odinger cat states. The size of the cat states can be inferred from the frequency of the interference fringes near the origin of phase space. Modulated Gaussian fits (solid blue lines) yield cat sizes of (\textbf{a}) $S=128 \pm 13$ photons (error bars are smaller than the data points) and (\textbf{b}) $S=1024 \pm 52$ photons. The dashed black lines show a rescaled Wigner distribution of the vacuum state for comparison. The Wigner distributions were obtained by measuring the expectation values of the displaced parity, omitting the $\pi/2$ normalization constant. The imperfect visibility of the interference fringes in (a) is largely due to transmon dephasing during state preparation and tomography, both of which take a time $\sim \SI{12}{\micro\second}$. In (b), the visibility is further reduced due to a variety of factors, such as the increased photon loss probability and spurious nonlinear effects as photon numbers approach $\bar{n}_\textrm{crit}^e$ (see Appendix \ref{Sec:Error_budget}). The simulated Kerr anharmonicity \cite{Kirchmair2013ObservationEffect} of the cavity $K_c/2\pi=3.6$ Hz is too small to substantially distort the cat states in our experiments (see Appendix \ref{Sec:Cavity_Kerr_nonlinearity}).}
  \label{fig:Wigner1dFringes}
\end{figure}
 
The size of the cat state, defined as $S=|2\alpha|^2$, corresponds to the square distance in phase space between the superimposed coherent states in units of photons \cite{Deleglise2008ReconstructionDecoherence,Vlastakis2013DeterministicallyStates}. To characterize the prepared Schr\"odinger cat states, we measure the fringes of the Wigner distribution near the origin of phase space, which arise from interference between the two opposite-phase coherent states (see Fig. \ref{fig:Wigner1dFringes}). The oscillation rate of this interference pattern is $2\sqrt{S}$ (see Appendix \ref{Sec:Determining the size of Schrodinger cats}), allowing for a reliable determination of the cat state size. Using this approach, we were able to demonstrate Schr\"odinger cat states with sizes up to $S=1024\pm52$ photons. This cat size corresponds to an average photon number of $|\alpha|^2=256\pm13$, which represents a tenfold increase over previous demonstrations using superconducting resonators \cite{Vlastakis2013DeterministicallyStates} and a twofold enhancement compared to phononic cat states of a single ion \cite{Johnson2017UltrafastAtom}. Further increasing the cat size would bring its average photon population close to the critical photon number $\bar{n}^e_\textrm{crit}\approx\frac{K_q}{6\chi}= 579$, where the dispersive approximation ceases to hold \cite{Eickbusch2022FastQubit}.

 As we increase the size of the Schr\"odinger cat states, their susceptibility to decoherence due to photon loss also increases. Indeed, the coherence of the cat state, as defined by the visibility of the interference pattern, is expected to decay exponentially with a relaxation rate $T_d^{-1}=S/(2T_1^c)$ \cite{Brune1992ManipulationStates} (see Appendix \ref{Sec:decoherence_rate}). In Fig. \ref{fig:CatParityLifetime}, we experimentally confirm the expected linear dependence of the decoherence rate on the cat size. For the largest cat states with a size of $S=1024$ photons, we measure a coherence time of $T_d=\SI{54 \pm 10}{\micro\second}$. This is significantly longer than the minimum achievable gate times $T_\textrm{g}^\textrm{min} \sim 1 / \left(\sqrt{\bar{n}_\textrm{crit}^e}\chi \right) = \SI{0.2}{\micro\second}$ \cite{Eickbusch2022FastQubit}, which opens the possibility for high-fidelity gates on these cat states.

\begin{figure}[t!]
  \includegraphics[width=0.45\textwidth]{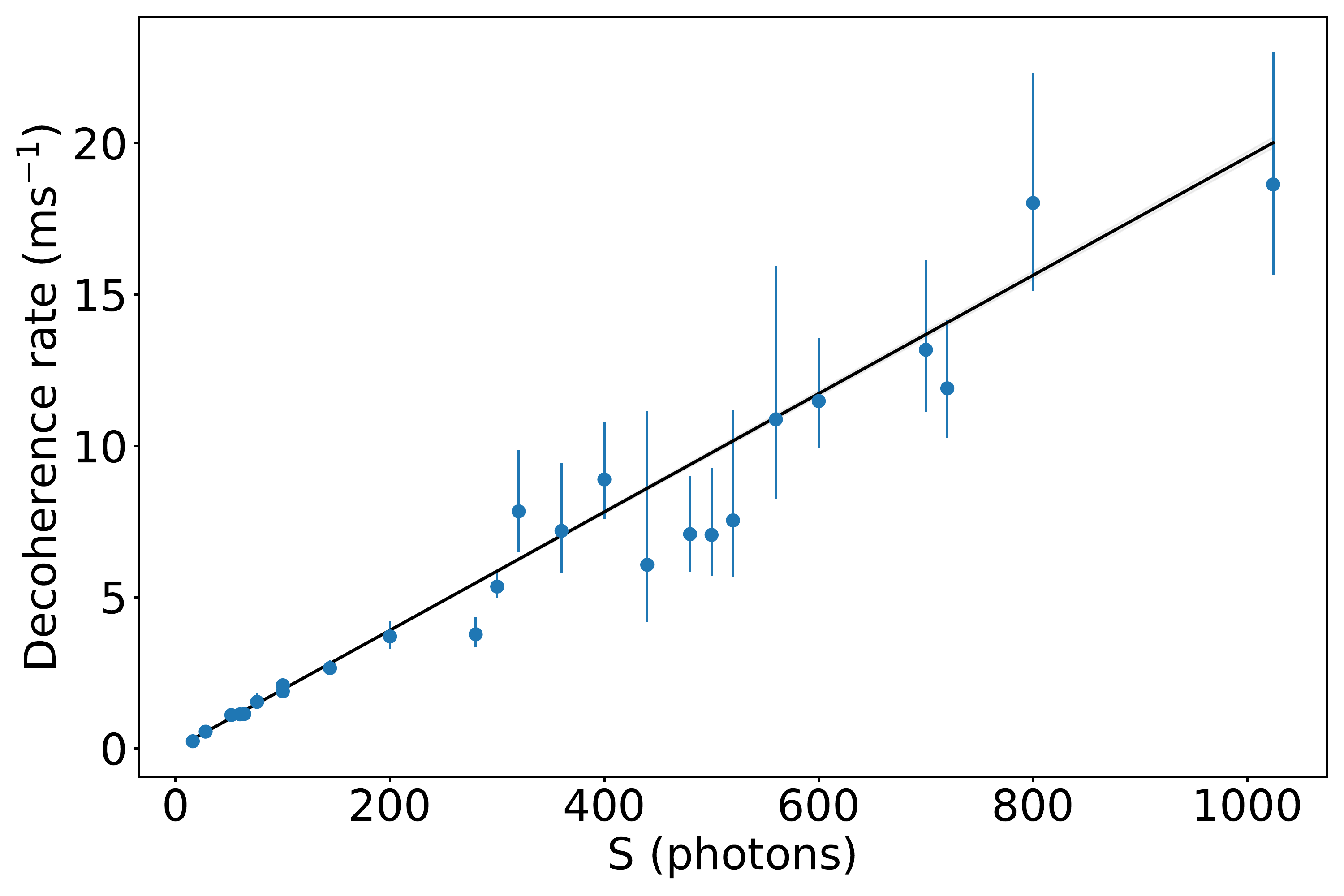}
  \vspace{-4pt}
  \caption{Decoherence rate $T_d^{-1}$ of Schr\"odinger cat states as a function of their size $S$. Each data point corresponds to the exponential decay rate of the central interference peak. The solid line shows the expected linear dependence on cat size $T_d^{-1}=S/(2T_1^c)$ without fitting parameters.}
  \label{fig:CatParityLifetime}
\end{figure}
 
\hfill

\section{Conclusion}
In this work, we introduced a novel superconducting microwave cavity whose single-photon qubit coherence times exceed the current state of the art by over an order of magnitude. The demonstrated qubit features a gap of four orders of magnitude between the photon lifetime and the gate time. This gap could enable high-fidelity quantum gates through the use of path-independence \cite{Ma2020Path-IndependentAncilla, Reinhold2020Error-correctedQubit} or erasure detection \cite{Barrett2010FaultErrors,Teoh2022Dual-railCavities,Kubica2022ErasureCircuits,Chou2023DemonstratingMeasurements,Levine2023DemonstratingTransmons}, mitigating the effect of transmon errors. Moreover, the long qubit coherence times can facilitate an in-depth study of subtle decoherence processes that are typically masked by photon loss \cite{Slichter2012Measurement-inducedNoise,Muller2019TowardsCircuits}.

We demonstrated the preparation of multiphoton states with large average photon numbers, while maintaining a low single-photon loss probability per operation. This combination paves the way for the implementation of bosonic quantum error correction protocols with the potential to significantly exceed the break-even point \cite{Ofek2016ExtendingCircuits,Sivak2023Real-timeBreak-even,Ni2023BeatingQubit}. Examples include rotation-symmetric codes \cite{Grimsmo2020QuantumCodes,Leviant2022QuantumChannel} and novel protocols for the Gottesman-Kitaev-Preskill (GKP) code \cite{Grimsmo2021QuantumCode,Siegele2023RobustError-correction,Campagne-Ibarcq2020QuantumOscillator,Sivak2023Real-timeBreak-even}.

Further improvements to the single-photon lifetime are within reach. For example, using a vacuum annealing step to remove surface oxides has been shown to significantly enhance the quality factor of niobium cavities \cite{Romanenko2020Three-DimensionalS,Posen2020UltralowCavities}.
These improved cavities can be coupled to superconducting qubits with longer coherence times \cite{Place2021NewMilliseconds,Wang2022TowardsMilliseconds, Somoroff2021MillisecondQubit}, reducing hybridization-induced photon loss and potentially enabling superconducting cavity qubits with coherence times approaching one second. Other avenues for improvement include active feedback cooling of the transmon \cite{Riste2012FeedbackMeasurement} to mitigate transmon-induced cavity dephasing. Furthermore, the fidelity of cavity state preparation and measurement can be substantially increased by using fast conditional displacements and qubit echo pulses \cite{Eickbusch2022FastQubit,Diringer2023ConditionalQubit}. 
To move towards the goal of realizing large-scale quantum computers, microwave cavities must be microfabricated into 3D integrated circuits \cite{Brecht2016MultilayerComputing,Lei2020HighBonding}. The insights gained from this work can be leveraged to improve the performance of these systems. 
\vspace{15 pt}
\begin{acknowledgments}
 \vspace{-5 pt}
We thank Benny Pasmantirer, Michael Rappaport, Haim Sade, and Zeng Hui for their contributions in designing
the experimental system. The cavity was manufactured by RI Research Instruments GmbH under the supervision of Peter vom Stein. We acknowledge input from David Schuster and Andrew Oriani on cavity-etching methods, and comments on the paper from Philippe Campagne-Ibarcq, Alec Eickbusch, and Shay Hacohen-Gourgy. We acknowledge financial support from the Israel Science Foundation ISF Quantum Science and Technologies Grant 963/19 and the European Research Council Starting Investigator Grant Q-CIRC 134847. S.R. is the incumbent of the Rabbi Dr Roger Herst Career Development
Chair.
\end{acknowledgments}


\appendix

\section{Hamiltonian and Lindbladian parameters}
\label{Sec:HamiltonianParameters}

The Hamiltonian of the full system, comprised of the cavity, the transmon and the readout resonator, can be expressed as
\begin{align}
    \label{Eq:Hamiltonian}
    \hat{\mathcal{H}} / \hbar &= \omega_c \hat{c} ^{\dagger} \hat{c} + \omega_q \hat{q} ^{\dagger}\hat{q} +\omega_r \hat{r} ^{\dagger}\hat{r}  \nonumber \\ 
    & - \frac{K_c}{2} \hat{c} ^{\dagger2} \hat{c}^2 - \frac{K_q}{2} \hat{q} ^{\dagger2} \hat{q}^2 - \frac{K_r}{2} \hat{r} ^{\dagger2} \hat{r}^2 \nonumber \\
    &- \chi \hat{c} ^{\dagger}\hat{c} \hat{q} ^{\dagger}\hat{q}  
    - \chi_{qr} \hat{r} ^{\dagger}\hat{r} \hat{q} ^{\dagger}\hat{q} - \chi_{cr} \hat{r} ^{\dagger}\hat{r} \hat{c} ^{\dagger}\hat{c},
\end{align}
where $\hat{c}$, $\hat{q}$, $\hat{r}$ are the annihilation operators of the cavity, transmon, and readout resonator, respectively.
The parameters for this Hamiltonian, along with relevant coherence properties, are detailed in Table \ref{Tab:HamiltonianParameters}.\\

\begin{table*}[t]
    \begin{tabular}{cll}
        \toprule
        \textbf{parameter} & \ \textbf{description} & \textbf{value} \\
        \hline \\
        $\omega_c$ & cavity resonance frequency & $2 \pi \times 4.301 \,$GHz \\
        $K_c$ & cavity self-Kerr & $2 \pi \times 3.6 \,$Hz$^{(*)}$\\
        $T_1^c$ & cavity lifetime & $25.6 \,$ms \\
        $T_2^c$ & cavity coherence time & $34 \,$ms \\
        $\Bar{n}^c_\text{th}$ & cavity average thermal population & $ < 0.5 \,$\% \\
        $\omega_q$ & transmon resonance frequency & $ 2 \pi \times 3.099 \,$GHz \\
        $K_q$ & transmon anharmonicity & $2 \pi \times 146 \,$MHz \\
        $T_1^q$ & transmon lifetime & $\SI{110}{\micro\second}$ \\
        $T_2^q$ & transmon coherence time & $\SI{16}{\micro\second}$ \\
        $T_{2 \! \hspace{0.8pt} E}^q$ & transmon Hahn-echo coherence time & $\SI{80}{\micro\second}$ \\
        $\Bar{n}^q_\text{th}$ & transmon average thermal population & $0.12 \,$\% \\
        $T_1^f$ & transmon $\ket{f}$-state lifetime & $\SI{50}{\micro\second}$ \\
        $T_2^{g \! \hspace{0.8pt} f}$ & transmon $(\ket{g}+\ket{f})$-superposition coherence time & $\SI{45}{\micro\second}$ \\
        $\chi$ & transmon-cavity dispersive shift & $2 \pi \times 42 \,$kHz \\
        $\omega_r$ & readout resonator resonance frequency & $2 \pi \times 7.889 \,$GHz \\
        $K_r$ & readout resonator self-Kerr & $2 \pi \times 2.3 \,$kHz$^{(*)}$\\
        $T_1^r$ & readout resonator lifetime & $\SI{0.38}{\micro\second}$\\
        $\chi_{qr}$ & transmon-readout dispersive shift & $2 \pi \times 1.3 \,$MHz \\
        $\chi_{cr}$ & cavity-readout dispersive shift & $2 \pi \times 0.2 \,$kHz$^{(*)}$ \\
        \hline \hline
    \end{tabular}
    \caption{System parameters and their respective values, cf. Eq. \eqref{Eq:Hamiltonian}. The table also includes relevant coherence times of the system. Values obtained through simulation \cite{Minev2021Energy-participationCircuits}, rather than direct measurement, are indicated with an asterisk.}
    \label{Tab:HamiltonianParameters}
\end{table*}

\begin{table*}[t]
\begin{tabular}{l @{\hskip 0.5cm} l @{\hskip 0.5cm} l}
\toprule
\textbf{description} & \ \textbf{Hamiltonian term} & \textbf{drive frequency}\\
\hline  & \\[-2ex]
cavity displacement drive & $\frac{\varepsilon_c}{2} \hat{c}^{\dagger}$ &  $\omega_c$\\& \\[-2ex]
readout measurement drive & $\frac{\varepsilon_r}{2} \hat{r}^{\dagger}$ & $\omega_r$\\& \\[-2ex]
transmon drive & $\frac{\varepsilon_q}{2}  \hat{q}^{\dagger}$ &  $\omega_q \ , \ \omega_q - K_q$\\& \\[-2ex]
cavity-transmon sideband drive & $\frac{\Omega}{2\sqrt{2}}\hat{q}^{2}\hat{c}^{\dagger}$ & $2\omega_q - K_q - \omega_c$\\& \\[-2ex]
cavity reset via readout resonator & $\frac{\Omega_{cr}}{2} \hat{c} \hat{r}^{\dagger}$ & $(\omega_r - \omega_c) / 2$\\& \\[-2ex]
\hline \hline
\end{tabular}
\caption{Drives used in this work along with their corresponding frequencies. $\varepsilon$ denote the drive rates of the respective elements, whereas $\Omega$ represent the rates of parametric interactions. Note that the cavity reset drive rate $\Omega_{cr}$ is proportional to the drive amplitude squared, whereas the other drive rates are linear in the applied amplitude. Hermitian conjugates are not shown in the Hamiltonian terms.}
\label{Tab:Pulses}
\end{table*}

The various drives used throughout this work are summarized in Table \ref{Tab:Pulses}.
To encode a single-photon qubit in the cavity, we use a driven four-wave mixing interaction that converts two transmon excitations into a single cavity excitation and vice versa \cite{Zeytinoglu2015Microwave-inducedElectrodynamics}. We address this sideband interaction by driving at a frequency $2 \omega_q - K_q - \omega_c$. The  sideband interaction rate is $\Omega = 2\xi\sqrt{K_q \chi}$ \cite{Pechal2014Microwave-controlledElectrodynamics,Rosenblum2018ACavities,Eickbusch2022FastQubit}, with $\xi$ the driven transmon displacement. Consequently, we can encode the single-photon qubit within $\SI{1}{\micro\second}$, significantly shorter than the relevant coherence times ($T_2^{g \! \hspace{0.8pt} f}=\SI{45}{\micro\second}$, $T_1^f=\SI{50}{\micro\second}$), enabling state preparation fidelities of $\sim 98 \%$. On the other hand, encoding schemes that rely on the dispersive interaction \cite{Heeres2017ImplementingOscillator} require a duration on the order of $2 \pi / \chi \sim \SI{24}{\micro\second}$ with our parameters, which would result in low state preparation fidelities.\\

Allowing the cavity state to passively relax to the ground state after each experiment is impractical due to the long single-photon lifetime of our cavity and the large number of stored photons. For the largest encoded cat state with initial average photon number $\bar{n}_i=256$, decaying to a final average population of $\bar{n}_f=0.01$ requires a wait time of $T_1^c\ln{\left(\frac{\bar{n}_i}{\bar{n}_f}\right)}=0.26\,$s. To enhance the duty cycle of our experiments, we adopt an active cooling method \cite{Wang2016ABoxes}. This method involves a four-wave mixing beam-splitting interaction that swaps a single cavity excitation with a single excitation in the readout resonator using a pair of drive photons at a frequency $(\omega_r - \omega_c) / 2$. The beam splitting rate is given by $\Omega_{cr}=2|\xi_{cr}|^2\sqrt{ \chi_{qr} \chi}$ \cite{Pfaff2017ControlledMemory}. Since any excitation in the lossy readout resonator is rapidly lost to the environment, this drive effectively induces an artificial photon loss mechanism to the cavity with a rate $\kappa_c^{\text{driven}}=\Omega_{cr}^2/\kappa_r$, where $\kappa_r=1/T_1^r$ is the relaxation rate of the readout resonator. Using this approach, we decrease the single-photon decay time to $0.6 \,$ms, which is over an order of magnitude faster than the intrinsic decay time. Consequently, a single photon can be removed from the cavity within $\sim 2 \,$ms, whereas for large cat states, we apply the drive for $\sim 8 \,$ms.

\section{Analysis of cavity loss channels}
\label{Sec:CavityLossChannels}
Several photon loss mechanisms contribute to the total cavity single-photon decay rate $\kappa_c^\text{total} = 1/T_1^c$.
Throughout the design of the experimental setup, we used a range of mitigation strategies to minimize these loss mechanisms. 
The total photon loss rate of the cavity is the sum of all loss channels, including both intrinsic losses and those related to coupling with the environment via RF couplers, i.e.,
\begin{align}
    \kappa_c^\text{total} = \sum_{\substack{\text{loss} \\ \text{channels}}} \kappa_c ^\text{channel}.
\end{align}
We now discuss the various loss mechanisms, estimate their contribution to the total photon loss rate, and elaborate on the corresponding mitigation methods. A summary of these details can be found in Table \ref{Tab:LossChannelsSummary}.

The aggregate of these losses results in a total loss rate of $\kappa_c^\text{total} = (\SI{120}{ms})^{-1}$ in the presence of the transmon chip. We stress that this simulated estimate excludes factors such as surface roughness, potential microcracks in the material, or residual surface resistance due to contaminants. Any of these effects may be responsible for the lower-than-expected bare cavity lifetime, as well as for its deterioration over time (see Fig. \ref{fig:ringdown}).

\subsection{Oxides on the cavity surface}
\label{Sec:Oxides_on_the_cavity_surface}
Oxides on surfaces of superconducting cavities contain a high density of two-level system defects \cite{Muller2019TowardsCircuits}, potentially resulting in significant photon loss \cite{Romanenko2017UnderstandingAmplitudes}. To mitigate this loss mechanism, we design the cavity geometry to minimize the effect of the lossy surface on the fundamental TM$_\text{010}$ cavity mode.
The effect of the dielectric oxide layer is quantified by the fraction of capacitive energy stored in this layer, known as the participation ratio or filling factor \cite{Gao2008ExperimentalResonators}:
\begin{align}
\label{Eq:FillingFactor}
F= \dfrac{\int _{V_\text{ox}} \frac{1}{2}\epsilon_\text{ox} \left|\vec{E} \right|^2 dV}{\int _{V_\text{total}} \frac{1}{2}\epsilon_0 \left| \vec{E} \right| ^2 dV},
\end{align}
where $\epsilon_\text{ox} = 33 \epsilon_0$ is the oxide's dielectric constant. The loss rate due to this effect is then given by \cite{Muller2019TowardsCircuits}
\begin{align}
    \kappa_c^\text{ox} = \omega_c F \tan \! \hspace{0.8pt} \delta_\text{ox} \tanh \left( \dfrac{\hbar \omega_c}{2 k_B T}\right),
\end{align}
with $\tan \! \hspace{0.8pt} \delta_\text{ox}\lesssim 10^{-2}$ the oxide's loss tangent  \cite{Wang2015SurfaceQubits,Read2022PrecisionSensitivity,Kaiser2010MeasurementResonators,Heidler2021Non-MarkovianMilliseconds}. Since the cavity temperature $T$ was kept at $\simeq 10 \, $mK throughout this work, we can use the approximation $\tanh { \left( \hbar \omega_c / 2 k_B T\right)} \approx 1$.

\begin{figure*}[t]
\centering
\includegraphics[width=0.9\textwidth]{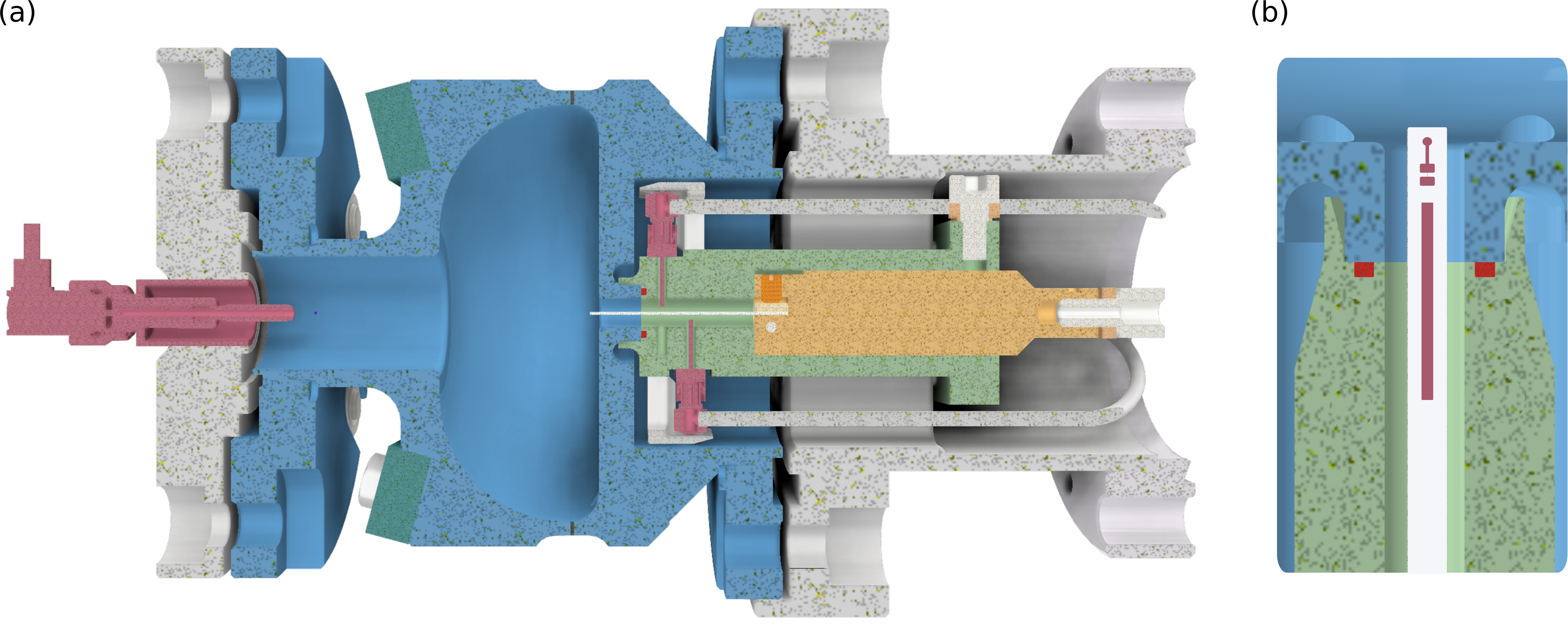}
\vspace{-5pt} \caption{ (\textbf{a}), Cross-section of the experimental setup. The niobium cavity is illustrated in blue, with the equatorial weld represented by black lines, and thermalization brackets in dark green. The cavity's inner diameter is \SI{59}{\mm} at the equatorial weld. The transmon chip (white) is held by a thermalized OFHC (Oxygen-Free High Conductivity) copper clamp (yellow) and enclosed within an aluminum housing (light green). An indium gasket (red) forms a superconducting seam between the cavity and the aluminum housing. The three RF couplers for driving the cavity, transmon, and stripline readout resonator, are shown in purple. (\textbf{b}), Top view of the sapphire chip (white) with aluminum (dark red) readout resonator and transmon. The transmon contains a small circular antenna, which slightly protrudes into the cavity volume.}
\label{Fig:SystemHalfSection}
\end{figure*}

Our design is similar to the elliptical TESLA cavity geometry \cite{Aune2000SuperconductingCavities}, which is frequently used for radio-frequency cavities in particle accelerators. Coupling a transmon to the maximum electric field of the fundamental mode in a TESLA cavity requires the chip to extend to the center of the cavity. This can exacerbate chip-induced losses, and can also introduce vibrations. Therefore, we modify the TESLA geometry by bisecting the cavity along the equator (see Fig. \ref{Fig:SystemHalfSection}). This preserves the mode structure through reflection symmetry, while allowing the transmon to access the maximum of the electric field with minimal protrusion of the chip into the cavity. The downside of this approach is that our cavity's filling factor is twice that of the full TESLA cavity ($F=1.4 \times 10^{-8}$ vs. $F = 0.7 \times 10 ^{-8}$), as determined by finite element simulations using a $5 \, $nm oxide layer thickness \cite{Romanenko2017UnderstandingAmplitudes}. However, our cavity's filling factor is five times lower than that of the coaxial stub cavity \cite{Reagor2016QuantumQED,Read2022PrecisionSensitivity} ($F = 7.6 \times 10 ^{-8}$), which is optimized for maximum coupling between the transmon and cavity modes. Focusing on achieving a low filling factor instead of maximizing the transmon-cavity coupling has led to a significant enhancement in the performance of our quantum memory. We estimate the loss rate attributed to this channel to be $\kappa_c^\text{ox} / 2 \pi \lesssim 0.6\,$Hz. A more rigorous approach would involve monitoring the thickness of the oxide layer and the cavity's surface roughness after etching, both of which can have a significant impact on the filling factor.

\subsection{Inverse Purcell loss}
\label{Sec:Inverse_Purcell_loss}
To achieve universal control, the cavity must be coupled to a chip-based nonlinear element, such as a transmon. This coupling modifies the single-photon state in the cavity to $ \ket{g,1}+(g_0/\Delta)\ket{e,0}$, with $g_0/2\pi=15.3 \, $MHz and $\Delta/2\pi=1.2\,$GHz denoting the cavity-transmon coupling rate and detuning, respectively. This modification inevitably introduces another photon loss channel to the cavity, known as the inverse Purcell effect \cite{Reagor2016QuantumQED}. The inverse Purcell loss is often approximated by multiplying the energy participation of the cavity mode in the transmon $p_{e,0}=(g_0/\Delta)^2$ by the transmon decay rate $\Gamma_\downarrow^q=1/T_1^q$. However, a more comprehensive approach also takes into account $\Gamma_\phi^q$, the pure transmon dephasing rate. This is because transmon dephasing errors can be viewed as projective transmon measurements, which have a probability of $p_{e,0}$ of transferring the cavity excitation to the transmon. This excitation then quickly dissipates due to transmon relaxation. Following Ref. \cite{Blais2021CircuitElectrodynamics}, the inverse Purcell effect including both mechanisms can be expressed as:
\begin{align}
\kappa_c^{\textrm{purcell}} &= \left( \dfrac{g_0}{\Delta} \right) ^2 \Gamma_\downarrow^q + 2\left( \dfrac{g_0}{\Delta} \right) ^2 \Gamma_\phi^q \nonumber \\
&\approx  2\left( \dfrac{g_0}{\Delta} \right) ^2 \dfrac{1}{T_{2 \! \hspace{0.8pt} E}^q}\approx \dfrac{\chi}{K_q}\dfrac{1}{T_{2 \! \hspace{0.8pt} E}^q},
\end{align}
where the last approximation is valid for  $K_q \ll \Delta$. In this expression, we use the Hahn-echo coherence time $T_{2 \! \hspace{0.8pt} E}^q$, since dephasing noise at low frequencies (below $\Delta$) does not result in excitation transfers to the transmon. To mitigate the inverse Purcell loss, we use a low dispersive coupling of $\chi = 2 \pi \times 42 \,$kHz, yielding an estimated loss rate of $\kappa_c^{\textrm{purcell}}/ 2\pi=0.6\,$Hz. Since the coupling of the cavity to the readout resonator is mediated by the transmon, this approach likewise reduces the readout-induced Purcell loss. As previously mentioned, the impact of reduced dispersive interaction rates can be partially offset by the use of strongly driven parametric interactions \cite{Rosenblum2018ACavities,Diringer2023ConditionalQubit,Eickbusch2022FastQubit}.

\subsection{Seam loss}
The two parts of the cavity are electron-beam welded to prevent seam loss. However, there is a seam between the cavity and the chip housing (see Fig. \ref{Fig:SystemHalfSection}). We minimize the contribution of the seam to the cavity photon loss by placing it at a depth of $6 \,$mm in a $2$-mm radius waveguide with cutoff frequency $57 \,$GHz. As a result, the magnetic field of the cavity mode is attenuated by three orders of magnitude, strongly suppressing currents through the seam. In addition, we use an indium gasket, which is compressed during integration of the chip. The resulting loss rate due to this seam can be expressed as the ratio between the seam conductance per unit length $g_{\text{seam}}$ and the seam admittance per unit length $y_{\text{seam}}$ \cite{Brecht2015DemonstrationCavities}:
\begin{align}
\kappa_c^{\text{seam}} = \omega_c \dfrac{y_{\text{seam}} }{g_{\text{seam}}}.
\end{align}
Using a simulated seam admittance of $y_{\text{seam}} = 3.3 \times 10^{-7} /\Omega$m and a seam conductance of $g_{\text{seam}} \sim 10^{6} /\Omega$m  \cite{Brecht2017MicromachinedCircuits}, we estimate a seam loss of $\kappa_c^\text{seam} / 2 \pi \sim 7.5 \times 10^{-4} \,$Hz.

\subsection{Conductive loss}
\label{Sec:conductive_loss}
Superconducting cavities exhibit finite surface resistance $R_\text{s}$, which causes conductive energy loss. The loss rate can be expressed as \cite{Padamsee2009RFApplications}
\begin{align}
    \kappa_c^{\text{cond}} = \omega_c \dfrac{R_\text{s}}{G}.
\end{align}
Here, $G$ is the geometry factor of the cavity, representing the ratio between the total electromagnetic energy in the cavity and the magnetic energy on its surface. The resistance due to thermal quasiparticles, as predicted by BCS theory, is exponentially suppressed as the temperature decreases and becomes negligible at \SI{10}{\milli\kelvin}. However, a residual surface resistance $R_\text{res}$ remains even as the temperature approaches zero (see Ref. \cite{Padamsee2009RFApplications} section §3.4).
Several mechanisms contribute to this residual surface resistance, such as trapped magnetic vortices. In this section, we detail our methods for reducing ambient magnetic fields and establish an upper bound on the residual surface loss caused by other potential imperfections like surface contaminants.

Our novel cavity design features a geometry factor that is higher than the commonly used stub cavity (\SI{210}{\ohm} vs. \SI{48}{\ohm}), making it less susceptible to residual surface resistance. This improvement, combined with the reduced filling factor (see Appendix \ref{Sec:Oxides_on_the_cavity_surface}), highlights the advantage of our design.

Experimental studies show that during cooldown, ambient magnetic fields can trap flux vortices, which introduce surface resistance \cite{Padamsee2009RFApplications}. For our cavity frequency, the surface resistance with an ambient magnetic field of $1 \,$mG is $R_\text{mag} \sim 2 \, \text{n}\Omega$.\\
To attenuate the ambient magnetic field during cooldown, we use two magnetic shields. The first shield, placed at room temperature, reduces Earth's magnetic field ($\sim 500 \,$mG) by two orders of magnitude. In addition, a $1\,$mm-thick Amumetal 4K shield encapsulates our cavity. This shield further reduces the strength of Earth's magnetic field, as well as stray magnetic fields from components inside the cryostat. Simulations indicate that this shield attenuates the magnetic field by three orders of magnitude.
Hence, the expected loss due to Earth's magnetic field is $\kappa_c^{\text{mag}} / 2\pi \sim 2 \times 10^{-4} \,$Hz.

Using the maximum quality factor measured in our cavity after etching ($Q_0=3\times10^9$), we can establish an upper bound for the residual surface resistance of $R_\text{res} < G / Q_0=\SI{70}{\nano\ohm}$. This upper bound is approximately an order of magnitude lower than the surface resistance observed in etched high-purity aluminum cavities \cite{Lei2023MicrowaveResonators}.

\subsection{Losses induced by the sapphire chip}
The presence of the sapphire chip in the cavity mode leads to bulk dielectric loss. To estimate the corresponding loss rate, we simulate the bulk participation ratio, yielding $p_\textrm{bulk}=1.0 \times 10^{-4}$, and use the experimentally determined \cite{Read2022PrecisionSensitivity} sapphire loss tangent $\tan \! \hspace{0.8pt}\delta_\text{bulk} = 6 \times 10^{-8}$. We can express the bulk loss rate as $\kappa_c^\text{bulk}=\omega_c p_\textrm{bulk} \tan \! \hspace{0.8pt}\delta_\text{bulk} $, yielding  $\kappa_c^\text{bulk}/ 2 \pi = 2.7 \times 10 ^{-2} \,$Hz.\\
In addition to bulk loss,  surface contaminants on the chip have been identified as a limiting factor for the lifetimes of chip-based  qubits \cite{Wang2015SurfaceQubits}. We estimate their effect on the cavity's lifetime using a similar approach, considering an estimated thickness of $3 \,$nm and the following loss tangents for metal-air, metal-sapphire and sapphire-air interfaces: $\tan \! \hspace{0.8pt}\delta_\text{MA}=2.1 \times 10^{-2}$, $\tan \! \hspace{0.8pt}\delta_\text{MS}=2.6 \times 10^{-3}$ and $\tan \! \hspace{0.8pt}\delta_\text{SA}=2.2 \times 10^{-3}$ \cite{Wang2015SurfaceQubits}. Using simulations to determine the respective participation ratios, we estimate the total loss rate due to these interfaces as $\kappa_c^\text{surface} / 2 \pi = 2.9 \times 10 ^{-2} \,$Hz.

The low photon loss rates introduced by the chip can be attributed to their small energy participation ratios in the cavity mode. This was achieved by minimizing the chip penetration into the cavity ($\sim 1$ mm), while still allowing for sufficiently large dispersive coupling between the transmon and the cavity.

\begin{table*}[t]
    \centering
    \begin{tabular}{|l|l|c|c|}
        \hline
        \textbf{Loss channel} & \textbf{Main mitigation methods} & \textbf{Loss rate} & \textbf{Lifetime} \\ && $\bm{\kappa_{c} / 2 \pi}$\textbf{ (Hz)}&  $\bm{T_1^c}$\textbf{ (s)} \\ \hline \hline
        Niobium surface oxides & 1. Filling-factor reduction & $6.0 \times 10 ^{-1}$ & $0.26$ \\ & 2. Chemical etching & & \\ \hline
        Inverse Purcell loss & 1. Weak cavity-transmon coupling & $5.7 \times 10 ^{-1}$ & $0.28$ \\ & 2. High-coherence transmon & & \\
        \hline
        Seam loss & 1. Seam located in narrow waveguide & $7.5 \times 10 ^{-4}$ & $211$ \\ & 2. Indium gasket & & \\
        \hline
        Sapphire bulk loss & Minor protrusion of the chip into the cavity & $2.7 \times 10^{-2}$ & $5.9$ \\ \hline
        Chip surface losses & Minor protrusion of the chip into the cavity & $2.9 \times 10^{-2}$ & $5.6$\\ \hline
        Magnetic vortices & Two magnetic shields & $2.0 \times 10^{-4}$ & $777$\\ \hline
        External coupling & RF ports undercoupled to cavity mode & $9.6 \times 10^{-2}$ & $1.7$\\ \hline
    \end{tabular}
    \caption{Summary of the loss channels and mitigation strategies. The values were obtained using finite-element numerical simulations, as detailed in the corresponding sections. We only considered the transmon and readout resonator couplers for the external coupling channel, as the cavity coupler was disconnected during the insertion of the transmon chip.}
    \label{Tab:LossChannelsSummary}
\end{table*}

\section{Cavity manufacturing and surface preparation}
\label{Sec:Cavity manufacturing and surface preparation}

The cavity is composed of two parts (see Fig. \ref{Fig:SystemHalfSection}): a flat section and a half-elliptical section, both of which were CNC machined from high-purity niobium with a residual resistivity ratio (RRR) of $\gtrsim 300$. The two parts were then joined using an equatorial electron-beam welding process. To eliminate damaged surface layers and oxides, the cavity underwent a $\SI{20}{\micro\meter}$ etch using buffered chemical polishing (BCP) with a $\text{1:1:2}$ $\text{HF:HNO}_3\text{:H}_3\text{PO}_4$ solution. This $15$-minute process required the cavity to be cooled to below  \SI{12}{\celsius} to maintain an etching rate of $\SI{1.35}{\micro\meter}$/min. Afterward, the cavity was thoroughly cleaned with a high-pressure water rinse.\\

Prior to cooling down, we conducted a second $\SI{30}{\micro\meter}$ BCP etch using a  $\text{1:1:2}$ $\text{HF:HNO}_3\text{:H}_2\text{O}$ solution \cite{Oriani2022MultimodalElectrodynamics} to remove any regrown oxides. This $5$-minute process required cooling the cavity to below \SI{10}{\celsius} to achieve an etching rate of $\SI{6}{\micro\meter}$/min. To minimize oxide regrowth, the cavity was cooled down in a dilution refrigerator within one hour of the final etch.\\ 

\section{Cavity dephasing induced by transmon excitations}
\label{Sec:ThermalPopulationDephasing}

The dominant source of dephasing in our cavity is the noisy transmon.
When the transmon is thermally excited from $\ket{g}$ to $\ket{e}$, it gains information on the cavity photon population, thereby inducing cavity decoherence. This behavior can be modeled as \cite{Rigetti2012SuperconductingMs}

\begin{align}
    \Gamma_\phi^\text{th} = \frac{\Gamma_\downarrow^q}{2} \Re \left( \sqrt{\left ( 1 + \frac{i \chi}{\Gamma_\downarrow^q}\right) ^2 + \frac{4 i  \chi \bar{n} ^q _{\text{th}}}{\Gamma_\downarrow^q}} - 1 \right),
\label{Eq:T_phi_thermal}
\end{align}
where $\Gamma_\downarrow^q=1/T_1^{q}$ is the transmon relaxation rate and $\bar{n} ^q _{\text{th}}$ is the average thermal population of the transmon. In the main text, the average time between transmon excitations $T _\uparrow ^q \approx T_1^q /\bar{n}_\textrm{th}^q$ is used for the transmon-induced dephasing time. This expression can be derived from Eq. \eqref{Eq:T_phi_thermal} in the limit $\chi/\Gamma_\downarrow^q \gg 1$.

\begin{figure*}[t]
\centering
\includegraphics[width=\textwidth]{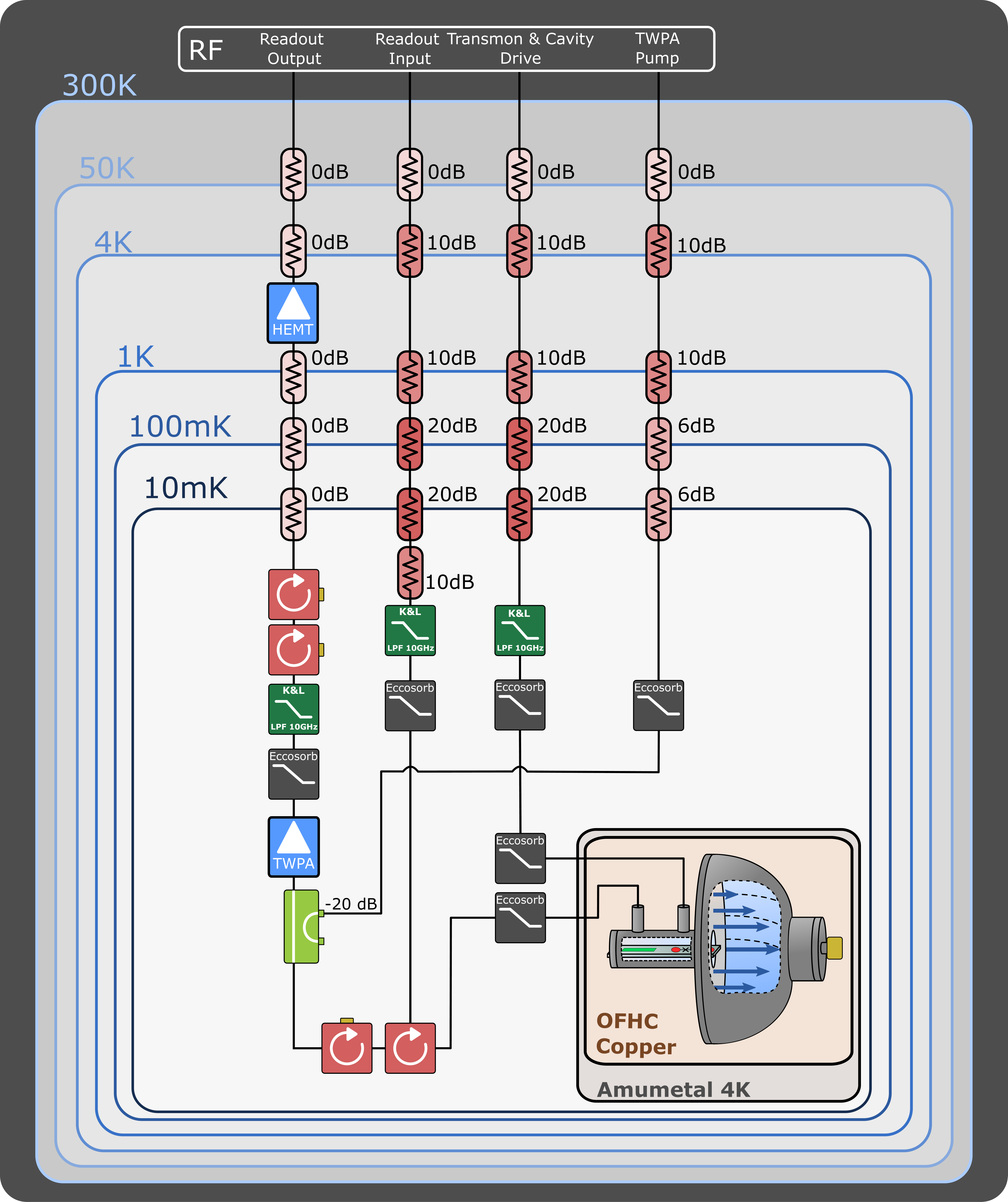}
\caption{Wiring diagram of the cryogenic microwave setup. The readout signal is first amplified using a Traveling Wave Parametric Amplifier (TWPA) from Silent Waves, followed by a HEMT amplifier from LNF at the 4K stage. The Eccosorb filters and the cryogenic directional coupler were supplied by Quantum Microwave. The control pulses were generated using Quantum Machines' OPX system before being up-converted to the system frequencies and sent into the fridge. For the classical ring-down experiment of Fig. \ref{fig:ringdown} in the main manuscript, we connected the readout input and output lines to the cavity RF pin. For the remainder of the experiments described in the manuscript, the cavity RF pin was disconnected, and instead, cavity driving was applied through the transmon drive line.}
\label{Fig:SetupDiagram}
\end{figure*}

Eq. \eqref{Eq:T_phi_thermal} highlights the importance of minimizing the thermal transmon population for achieving extended cavity coherence times. To accomplish this, we use several strategies. First, we use multiple thermal straps to directly thermalize both the OFHC copper chip clamp and the cavity to the cryostat's mixing chamber plate, ensuring effective cooling. To protect the transmon from high-frequency radiation, we place \SI{10}{\GHz} low-pass filters and Eccosorb infrared filters on all lines. We position the Eccosorb filters as close as possible to the RF coupling ports (see Fig. \ref{Fig:SetupDiagram}).  In addition, we use room-temperature RF switches to prevent near-resonant noise generated by control electronics from reaching the transmon during idling times. Furthermore, we encapsulate the entire experimental system with a thermalized OFHC copper shield, creating a cold and light-tight environment. To further reduce stray infrared radiation, we apply Eccosorb paint near the undercoupled cavity RF pin. 

As highlighted in Table \ref{Tab:CooldownsParameters}, the implementation of these measures has led to a reduction in the thermal transmon population by two orders of magnitude, which in turn resulted in a dramatic increase in cavity coherence time.

\setlength{\tabcolsep}{10pt}
\begin{table*}[t]
\begin{tabular}{lccccc}
\hline\hline & \\[-2ex]

 & $\bm{\bar{n} ^q _{\textbf{th}}}$ \textbf{(\%)} & $\bm{T_2^c}$ \textbf{(ms)} & $T_1^c$ (ms) & $\Gamma_\downarrow^q/2\pi$ (kHz) & $\chi/2\pi$ (kHz) \\ [0.5ex] 
\hline  & \\[-2ex]
Cooldown I     &     \textbf{7.2}                   &   \textbf{1.2}      &    25.7    &      1.88       &     197.0 \\
Cooldown II    &     \textbf{2.9}                   &    \textbf{1.2}    &    22.3    &       3.75      &     36.1 \\
Cooldown III   &     \textbf{0.1}                   &   \textbf{34.0}     &   25.6    &      1.45       &     42.0 \\
\hline\hline
\end{tabular}
\caption{Progression of cavity coherence times across multiple cooldown cycles. The results discussed in this work were acquired during Cooldown III, when all thermalization measures described in Appendix \ref{Sec:ThermalPopulationDephasing} were implemented. As evident from the data, these efforts significantly reduced the thermal transmon population, leading to improved cavity coherence times. By using Eq. \eqref{Eq:T_phi_thermal} and the parameters provided for each cooldown, the measured coherence times are found to be consistent with expectations. Furthermore, we note that the single-photon lifetime of the cavity remains consistent across various cooldowns, even though each cooldown involves a different transmon chip and is preceded by a cavity etch process.}
\label{Tab:CooldownsParameters}
\end{table*}

\section{Transmon chip integration}
\label{Sec:Transmon chip integration}

The transmon chip was diced from a c-plane EFG sapphire wafer with a thickness of $\SI{430}{\micro\meter}$. The aluminum transmon and stripline readout resonator were deposited in a single step using shadow evaporation. For assembly, the chip was inserted into a cylindrical aluminum housing using an oxygen-free high-conductivity (OFHC) copper clamp. This aluminum package was then connected to the narrow waveguide at the center of the cavity using an indium gasket, ensuring a low-loss seam. 

\section{Determining the size of Schr\"odinger cats}
\label{Sec:Determining the size of Schrodinger cats}
The sizes of the Schr\"odinger cats in Fig. \ref{fig:Wigner1dFringes} were determined by measuring a one-dimensional cut along the imaginary axis of the Wigner tomogram of the cavity. The obtained data were fitted to a modulated Gaussian function $\frac{A}{\sigma\sqrt{2\pi}}e^{-(x-\mu)^2/2\sigma^2}\cdot\sin\left({fx}+\phi\right)$ \cite{Haroche2006ExploringPhotons}, where the displacement along the imaginary axis $x$ was scaled so that the measured Gaussian Wigner function for the vacuum state gives a standard deviation of $\sigma=1/2$. The cat size is then found using $S = f^2/4$, with $f$ the modulation frequency obtained from the fit. The amplitude $A$, center $\mu$, and phase $\phi$ of the modulated Gaussian are free parameters.

\section{Cat state decoherence rate as a function of its size}
\label{Sec:decoherence_rate}

In this section, we elaborate on the decoherence of Schr\"odinger cat states in the presence of photon loss. The coherence of a Schr\"{o}dinger cat state can be characterized by the visibility of interference fringes in its Wigner distribution. This can be conveniently quantified by analyzing the value of the Wigner function at the phase space origin, which evolves over time as \cite{Brune1992ManipulationStates,Haroche2006ExploringPhotons}

\begin{align}
    W_\pm( & \beta=0, \Delta t) = \dfrac{4}{\pi (1 \pm e^{-2 \Bar{n}})} \times \nonumber \\
    &\left \{ \exp [-2 \Bar{n} e^{-\Delta t/ T_1^c} ] \pm  \exp \left[ -2 \Bar{n} \left(1 - e^{-\Delta t/ T_1^c} \right) \right] \right \},
\end{align}
where $\pm$ denotes the initial parity of the cat state and $\Bar{n}$ represents the initial mean photon number.
In the limit of large Schr\"odinger cat states ($e^{-2\Bar{n}} \ll 1$) and short times ($\Delta{t} \ll T_1^c$) we obtain
\begin{align}
\label{Eq:CatStateDecoherenceRate}
    W_\pm(\beta=0, \Delta{t}) &\simeq \pm \dfrac{4}{\pi}  \exp \left[-2 \Bar{n} \left(1 - e^{-\Delta{t}/ T_1^c} \right) \right]  \nonumber \\
        &\simeq \pm \dfrac{4}{\pi}  e^{-2\Bar{n} \Delta t / T_1^c }  \nonumber\\ 
        &\Rightarrow T_d = \dfrac{T_1^c}{2 \Bar{n}} = \dfrac{2T_1^c}{S},
\end{align}
where $S=4 \Bar{n}$ is the size of the cat state. As stated in the main text, this expression indicates that the decoherence rate of a cat state depends linearly on its size.\\

\begin{figure}[h!]
\centering
\includegraphics[width=0.45\textwidth]{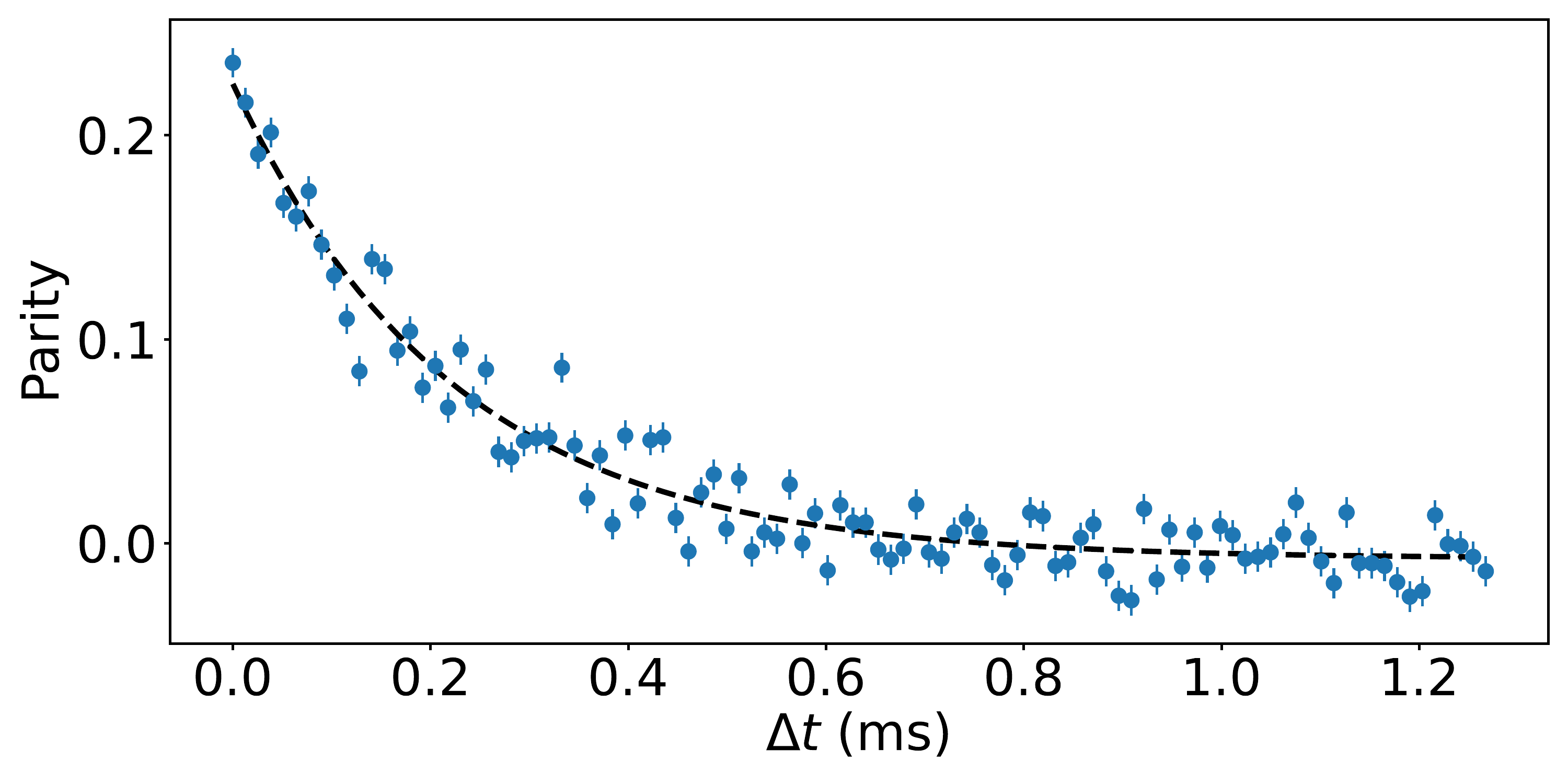}
\vspace{-4pt}
\caption{Measurement of decoherence rate of a cat state with size $S=200$ photons. The dashed line represents an exponential fit with a characteristic time of $T_d = \SI{222 \pm 13}{\micro\second}$.}
\label{Fig:decoherence_rate_measurement}
\end{figure}

To experimentally measure the decoherence rate, we use the fact that the Wigner distribution at the phase space origin is proportional to the average parity of the state. Therefore, we measure the average parity after varying time intervals, and extract the decoherence rate using an exponential fit. For example, Fig. \ref{Fig:decoherence_rate_measurement} shows the data used to extract the decoherence rate for $S=200$ photons, as presented in Fig. \ref{fig:CatParityLifetime} of the main text.

\section{Calibrating the transmon drive for parity measurements}
\label{Sec:Parity_measurement_with_detuned_drive}
To prepare a Schr\"odinger cat state in the cavity, we first displace the cavity state and then apply a parity measurement.  Depending on the parity measurement outcome, either an even or an odd cat state is obtained. 
However, when dealing with states containing a large average number of photons $\bar{n}$, the resonance frequency of the transmon shifts significantly, requiring the adjustment of the transmon drive frequency. For example, in the case of the ``chonk'' cat state with a size of $S=1024$ photons,  the transmon frequency shift is $\bar{n}\chi/2\pi \approx  10.8\,$MHz. However, only specific transmon drive frequencies can be used to implement a successful parity measurement \cite{Sun2014TrackingMeasurements}. To find these frequencies, we analyze the system's state during the free-evolution segment of the parity measurement. This state can be represented as $\ket{\psi (t)}=\ket{g}\ket{\alpha}+e^{-i(\omega_d - \omega_q -\chi \hat{n})t}\ket{e}\ket{\alpha}$, with the normalization constant omitted and with $\omega_d, \ \omega_q$ the drive and qubit bare frequency, respectively. For a time evolution of $T=\pi/\chi$ to correspond to a parity measurement, the drive detuning $\Delta \omega = \omega_d - \omega_q$ must satisfy  $\Delta \omega \cdot T = \pi k, \ k \in \mathbb{Z}$.

\begin{figure*}[t]
\centering
\includegraphics[width=\textwidth]{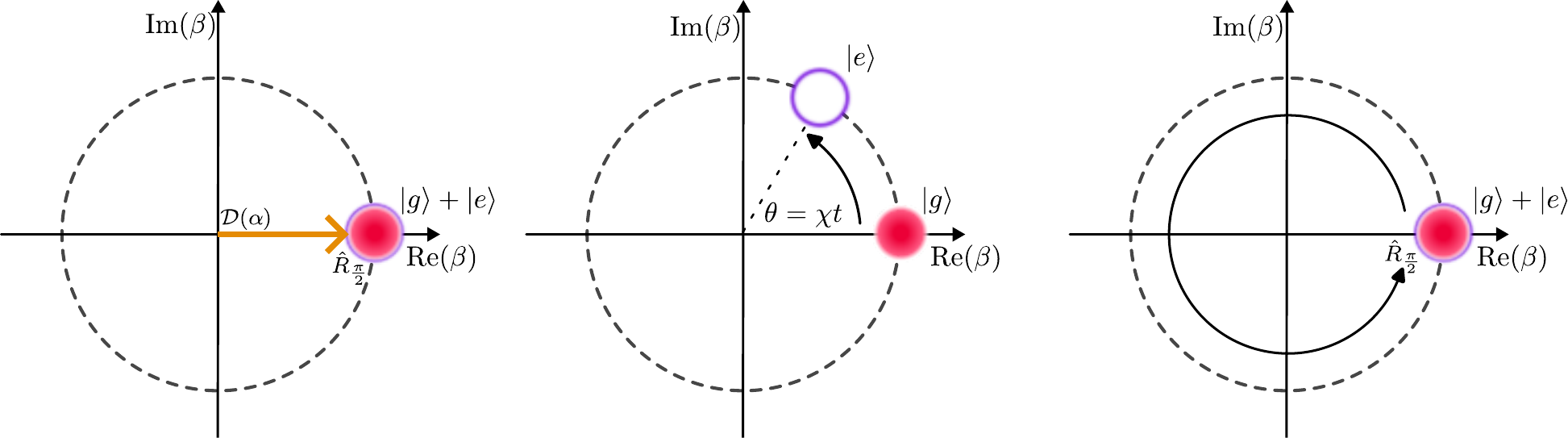}
\vspace{-4pt}
\caption{Schematic of transmon drive calibration for parity measurements. We first displace the cavity into the coherent state $\ket{\alpha}$ and initialize the transmon in $\ket{g} + \ket{e}$. The system is then allowed to freely evolve for a duration $2T=\frac{2\pi}{\chi}$, during which the coherent state associated with $\ket{e}$ accumulates a phase $\theta = 2\chi T=2\pi$. Finally, we apply a second $\pi/2$ transmon rotation, followed by a transmon measurement. Experimental data for this calibration procedure are shown in Fig. \ref{Fig:ParityCalibrationData}.}
\label{Fig:ParityCalibrationIllustartion}
\end{figure*}

\begin{figure}[h!]
\centering
\includegraphics[width=0.45\textwidth]{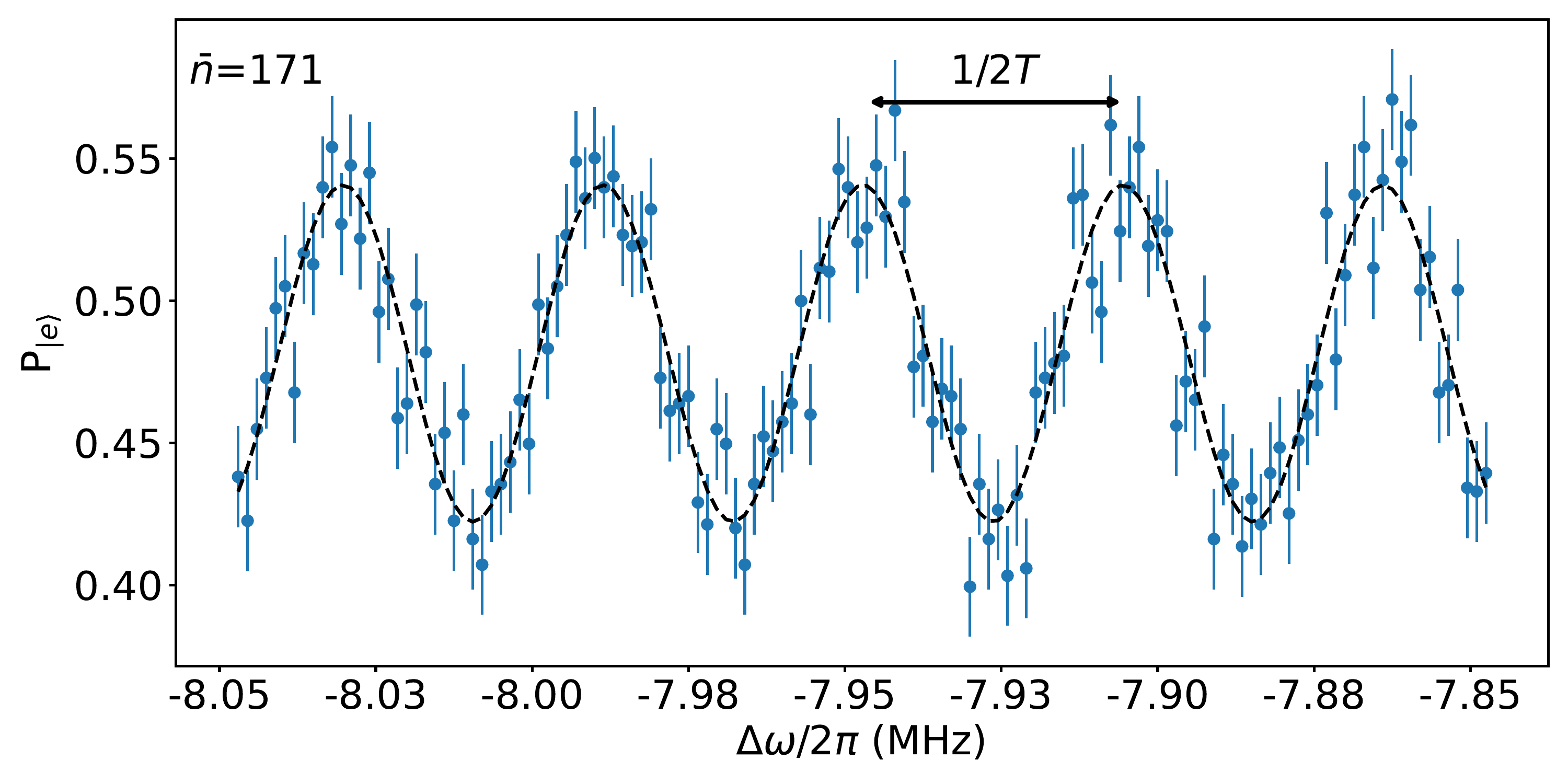}
\vspace{-4pt}
\caption{Transmon drive calibration experiment for parity measurements. This plot shows the probability of the transmon being in the excited state as a function of drive detuning from its bare resonance $\Delta\omega$, as detailed in Appendix \ref{Sec:Parity_measurement_with_detuned_drive}. The black dashed curve represents the fit to $A\cos(\Delta\omega/\nu+\phi) + B$. The fitted period $\nu$ is consistent with the period expected from Eq. \eqref{Eq:ParityCalibrationProbability}, i.e., $1/2T = \chi / 2 \pi$ . The signal visibility $A$ is low, primarily due to transmon decoherence, as explained in Appendix \ref{Sec:Error_budget}. The reduction in visibility is more pronounced in this experiment than in a parity measurement, since the state evolves for twice as long ($2\pi/\chi$ instead of $\pi/\chi$). These data were collected with a mean cavity photon number of $\bar{n}=171$ photons.}
\label{Fig:ParityCalibrationData}
\end{figure}

To calibrate the transmon drive frequency for each cavity displacement, we use the protocol depicted in Fig. \ref{Fig:ParityCalibrationIllustartion}. After displacing the cavity state, we apply an unconditional $\pi/2$ transmon  rotation. We then wait for a time $2T = 2 \pi/\chi= \SI{23.8}{\micro\second}$ for coherence revival, followed by an additional unconditional transmon $\pi/2$-pulse $\hat{R}_{\pi/2}$. The state of the system and the corresponding probability of measuring the transmon in $\ket{e}$ are then given by
\begin{align}
    \hat{R}_{\frac{\pi}{2}} \ket{\psi(2T)} =\frac{1}{2} \big[ &\ket{g} \left( 1-e^{-i\Delta\omega \cdot 2T} \right) \nonumber \\
    + &\ket{e} \left( 1+e^{-i\Delta\omega \cdot 2T} \right) \big] \ket{\alpha} \nonumber \\
      \Longrightarrow P_{\ket{e}} = \frac{1}{2} &+ \frac{1}{2} \cos(\Delta \omega \cdot 2T).
     \label{Eq:ParityCalibrationProbability}
\end{align}
The optimal transmon frequency corresponds to the maximum of the resulting cosine, which is closest to the mean transmon frequency shift (see Fig. \ref{Fig:ParityCalibrationData}).\\

\section{Error budget for cat state preparation and measurement}
\label{Sec:Error_budget}
\begin{figure}[h!]
\centering
\includegraphics[width=0.45\textwidth]{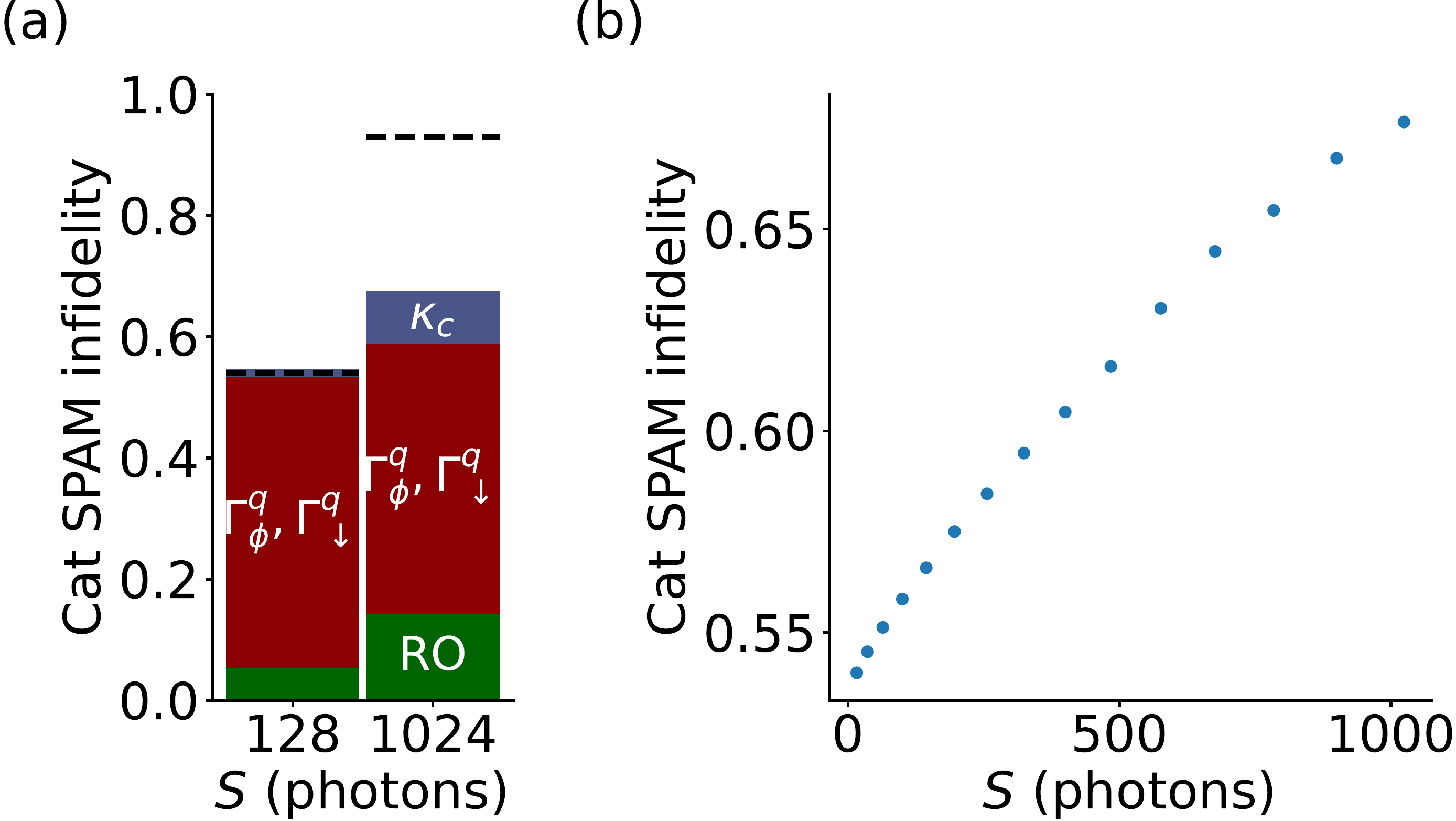}
\vspace{-4pt}
\caption{Error budget for cat state preparation and measurement (SPAM). (\textbf{a}), Simulation of the contribution of various SPAM error mechanisms to the interference fringe visibility for the cat states shown in Fig. \ref{fig:Wigner1dFringes} of the main text. RO refers to the experimentally determined transmon readout assignment infidelity in the presence of cat states. $\Gamma^q_{\phi}$ and $\Gamma^q_{\downarrow}$ denote transmon dephasing and decay, respectively, and $\kappa_c$ represents cavity photon loss. Dashed lines indicate the measured fringe visibility, which serves as an experimental measure of SPAM fidelity. While the simulations align with the observed fringe visibility for the small cat state, the fringe visibility of the large cat state is significantly lower than expected. (\textbf{b}), Simulated cat SPAM infidelity as a function of cat size. As the number of photons populating the cavity increases, the visibility of the cat fringes is reduced due to lower transmon measurement fidelity and a higher photon loss rate.}
\label{Fig: ErrorBudget}
\end{figure}

The visibility of the interference fringes for the cat states analyzed in this work is limited, as evident from Fig. \ref{fig:Wigner1dFringes} and the error bars in Fig. \ref{fig:CatParityLifetime} of the main text. This reduction in visibility is due to a combination of errors occurring during cat state preparation and measurement (SPAM). In this section, we explore the decoherence sources contributing to the reduced fringe visibility (see Fig. \ref{Fig: ErrorBudget}). 

The primary sources of decoherence in our system are transmon decay and dephasing (see Table \ref{Tab:HamiltonianParameters}). During parity measurements for state preparation and tomography, the transmon becomes entangled with the cavity state. Consequently, transmon errors can corrupt the cavity state or lead to inaccurate measurement outcomes. Within the dispersive approximation framework, the decrease in fringe visibility due to transmon decoherence is independent of the cavity photon number.

Cavity decay also contributes to cat state decoherence, as described in Appendix \ref{Sec:decoherence_rate}. Lost cavity photons carry information on the phase of the cavity state to the environment, decohering the entangled state and reducing SPAM fidelity. As larger cat states populate the cavity, information leakage increases, leading to greater decoherence due to photon loss.

The final step in both cat state preparation and tomography is to measure the transmon state. Therefore, transmon measurement infidelity also contributes to the decreased fringe visibility. We observed a reduction in readout fidelity when the cavity contains a high photon number. The readout assignment fidelity was $\mathcal{F}=0.95$ for cat states with few photons, degrading to $\mathcal{F}=0.86$ for the ``chonk'' cat state with size $S=1024$ photons. Further investigation is required to determine the mechanism behind this degradation in measurement fidelity.

To estimate the contribution of each error source to the reduction in fringe visibility, we conducted full quantum simulations \cite{Johansson2013QuTiPSystems} using the system Hamiltonian (Eq. \eqref{Eq:Hamiltonian}), excluding terms involving the readout resonator. The results are presented in Fig. \ref{Fig: ErrorBudget}. Fig. \ref{Fig: ErrorBudget}(a) shows that the observed fringe visibility for small cats is consistent with simulations. In contrast, for larger cats with $S=1024$, the observed fringe visibility is significantly lower than the simulated visibility including all the error sources mentioned above. Simulations using a simple model beyond the dispersive approximation \cite{Blais2021CircuitElectrodynamics} do not predict a substantial reduction in fringe visibility for $\bar{n}=256$ photons, despite being relatively close to the critical photon number $\bar{n}_\textrm{crit}^e\approx579$ photons. Potential factors contributing to the observed infidelity are heightened sensitivity to calibration errors for larger cat states and nonlinear effects not included in the model under consideration.

\section{Cavity Kerr nonlinearity}
\label{Sec:Cavity_Kerr_nonlinearity}
The coupling between the cavity and the transmon results in a nonlinear shift in the cavity frequency. This shifted frequency can be expressed semiclassically as $\omega_c(n)=\omega_c(n=0) - \frac{K_c}{2}n^2$ (cf. Eq. \eqref{Eq:Hamiltonian}). Here, $K_c\approx \chi^2/4K_q$ represents the cavity Kerr nonlinearity, with $K_c=2\pi \times 3.6\,$Hz for our cavity. The Kerr effect distorts coherent states by inducing a photon-number dependent angular velocity in phase-space, given by $\frac{\partial \omega_c(n)}{\partial n} = -K_c n$. A complete phase collapse of the coherent state occurs after a time $T_\textrm{col}=\frac{\pi}{2\sqrt{\bar{n}}K_c}$, when the difference in rotation angle across the $2\sqrt{\bar{n}}$ width of the photon number distribution reaches $\pi$ \cite{Kirchmair2013ObservationEffect,Haroche2006ExploringPhotons}. However, the distortion induced by the Kerr effect during the preparation and measurement of cat states is negligible. Indeed, even as $\bar{n}$ approaches the critical photon number $\bar{n}_\textrm{crit}^e\approx \frac{K_q}{6\chi}$, the duration of a parity measurement $T=\pi/\chi$ remains much shorter than the phase collapse time, since $\frac{T}{T_\textrm{col}}\approx \frac{1}{12\sqrt{\bar{n}_\textrm{crit}^e}}\ll 1$. For the largest cat state in our work, $T_\textrm{col}=4\,$ms. Not only is this duration longer than the preparation and measurement times, but it also exceeds the $\SI{54}{\micro\second}$ coherence time of the cat state due to photon loss by two orders of magnitude.

\clearpage

\bibliography{bibliography.bib}

\end{document}